\documentclass[a4paper]{article}
\usepackage{INTERSPEECH2018}
\usepackage{multirow}
\usepackage{graphicx}
\usepackage{soul}

\title{Deep neural networks for emotion recognition\\ combining audio and transcripts}
\name{Jaejin Cho$^1$,Raghavendra Pappagari$^1$,Purva Kulkarni$^2$,\\Jes\'{u}s Villalba$^1$,Yishay Carmiel$^2$,Najim Dehak$^1$}
\address{
  $^1$Center for Language Speech Processing, Johns Hopkins University, Baltimore, MD, USA\\
  $^2$IntelligentWire, Seattle, WA, USA}
\email{\{jcho52,rpappag1,jvillal7,ndehak3\}@jhu.edu,  \{pkulkarni,ycarmiel\}@intelligentwire.com}

\begin{document}

\maketitle
\begin{abstract}
In this paper, we propose to improve emotion recognition by combining acoustic information and conversation transcripts.
On the one hand, a LSTM network was used to detect emotion from acoustic features like f0, shimmer, jitter, MFCC, etc.
On the other hand, a multi-resolution CNN was used to detect emotion from word sequences. This CNN consists of several parallel
convolutions with different kernel sizes to exploit contextual information at different levels. A temporal pooling layer aggregates the
hidden representations of different words into a unique sequence level embedding, from which we computed the emotion posteriors.
We optimized a weighted sum of classification and verification losses. The verification loss tries to bring embeddings from
same emotions closer while separating embeddings from different emotions.
We also compared our CNN with state-of-the-art text-based hand-crafted features (e-vector).
We evaluated our approach on the USC-IEMOCAP dataset as well as
the dataset consisting of US English telephone speech. In the former, we used human-annotated transcripts while in the latter, we used ASR transcripts. The results showed fusing audio and transcript information improved unweighted accuracy by relative 24\% for IEMOCAP and relative 3.4\% for the telephone data compared to a single acoustic system. 
\end{abstract}

\section{Introduction}

Emotion recognition from speech has attracted attention because of its application in
human-computer interaction, affective learning systems, mental health analysis,
improvement of customer service, etc~\cite{picard2000affective}.
For example, tracking the user's emotional states during a call to a customer service
can help the agent to adapt his/her response to provide a better service. It can also be used
to evaluate the quality of the service provided by the agent.

Researchers have used different modalities to predict emotional states.
Computer Vision, speech processing and fusion of them are the most common modalities while there are also some works detecting emotion from transcripts~\cite{poria2017review}.
In this paper, we focus on the systems using either acoustic or textual information and the fusion of both,
all of which can be derived from speech.

In machine learning perspective,
the speech emotion recognition research has been done in mainly two directions: exploring emotion representative features~\cite{kim2013deep}
or building classifiers adapted to emotion recognition~\cite{rozgic2012svmensemble}. Nowadays, it is even possible to work on both using deep learning framework~\cite{mirsamadi2017rnnattention}.
The authors in~\cite{mirsamadi2017rnnattention} studied deep neural networks to generate feature vectors at frame-level from raw spectral representation,
aggregate the features over time into an utterance-level feature vector and classify it with softmax layer.
The systems were compared with SVM and DNN trained on hand-crafted features called low level descriptors (LLDs)
and statistical functions applied to them. In the experiments,
a LSTM network taking global temporal average pooling over a sequence of hidden layer vectors
showed significantly higher accuracy compared to other LSTM architectures, e.g.,
 taking the output of the LSTM at the last time step or the outputs over time frames without pooling.
Thus for our acoustic system, we used a LSTM network with a global temporal mean pooling layer with some variations in the LSTM structure.

Different from the above work, contextual LSTM networks were proposed in~\cite{poria2017context} where multiple utterance-level features are
fed to the networks to model dependencies between utterances within a video or session.
This method improved performance by 5-10\% compared to systems that do not consider the context information beyond utterance-level.
However, considering the context to such degree might not be available in some applications, e.g., real-time human-computer interaction.
We did not consider dependencies beyond utterance-level in this paper.

A number of acoustic and lexical features are explored in~\cite{jin2015speech}.
In the paper, performance of many differently combined fusion systems is reported.
In a pair-wise late fusion schemes, fusion of acoustic and lexical feature based systems showed
larger gains compared to other fusion systems, which we also explore in this paper.

In this paper, we study how an acoustic system improves when combined with a transcript based system.
The emotion recognizer from transcripts was based on a recently proposed multi-resolution CNN architecture~\cite{raghu2018mcnn}.
First, we compare our proposed text based CNN  with a SVM system based on emotion vector (e-vector)~\cite{jin2015speech}.
Then, we show how fusion of transcript based and acoustic system improve performance.
We experimented emotion recogntion task on IEMOCAP  and
call-center telephone conversations. For IEMOCAP, we used human annotated transcripts while in the call-center scenario we used ASR transcripts, which is more realistic.
The proposed CNN system outperformed a system trained on hand-crafted e-vector features.
Also, fusion results on the call center data indicates that adding information from ASR transcripts improves the performance of the acoustic LSTM system.

The organization of the paper is as follows. Section 2
explains the emotion recognition system based on acoustic features.
Section 3 describes the Multi-resolution CNN for emotion recognition from transcripts.
The IEMOCAP and call-center datasets used in our experiments are introduced in Section 4.
In Section 5, we explain the experimental setup and the results.
The results are compared to other systems proposed in previous works.
Finally Section 6 summarizes the paper and talks about future research directions.

\section{Acoustic emotion recognition with LSTM}

A speech utterance is composed of a sequence of feature frames. For this reason, we desire approaches that can model the temporal dependencies between frames. 
A natural choice is recurrent neural networks. They, however, fail to learn long-term dependencies due to the vanishing gradient problem. This led to the 
invention of LSTM~\cite{hochreiter1997lstm} and GRU~\cite{chung2014gru}  networks. These networks mitigate the vanishing gradient problem using more sophisticated structure that 
includes a memory cell. A set of non-linear functions, called gates, decides when to write or read data from the memory cell. 

LSTM networks are explored in several previous works in speech emotion recognition. 
The authors in~\cite{mirsamadi2017rnnattention} explored bidirectional neural networks (BLSTM) combined with several strategies 
and showed having a global average pooling layer in the network 
improves performance over other BLSTMs.  Another approach predicts the output by using only the last frame assuming the information in whole sequence is perfectly accumulated by LSTM memory cells.
However, though LSTMs reduce the vanishing gradient problem, they do not eliminate it completely. Thus, 
the information of the early frames still vanishes in layer time steps. In practice, LSTMs are able to consider only a few seconds of 
contextual information. For this reason, having a global temporal pooling helps. Thus in this paper, we 
used a LSTM with temporal pooling as our acoustic system. LSTMs were implemented using the Keras toolkit~\cite{chollet2015keras}.
The training objective function was categorical cross-entropy.

For the acoustic features used as inputs to the network, we used 
utterance-level sequences of 88 dimensional features from the extended Geneva Minimalistic Acoustic Parameter Set~\cite{eyben2016gemaps}. 
This set includes parameters such as pitch, jitter, shimmer, formants, MFCC, etc. called LLDs plus the statistical functions (mean, variance, min, max, etc.) 
applied to the LLDs over specified time sliding window. 
In this work, the frame size was set to 20ms with 10ms overlap and the statistical functions were applied over 60ms. 
The openSMILE~\cite{eyben2010opensmile} open-source software was used to extract the features.

\section{Text based emotion recognition with CNN}

To exploit information in transcripts, we used a multi-scale Convolutional Neural Network (CNN) framework, 
which showed state-of-the-art results on two datasets in topic identification task~\cite{raghu2018mcnn}. 
We expected that emotional utterance-level embeddings generated from the CNN can be used for emotion classification as document-level embeddings were for topic identification.

\begin{figure}[ht!]
  \centering
  \includegraphics[width=\linewidth]{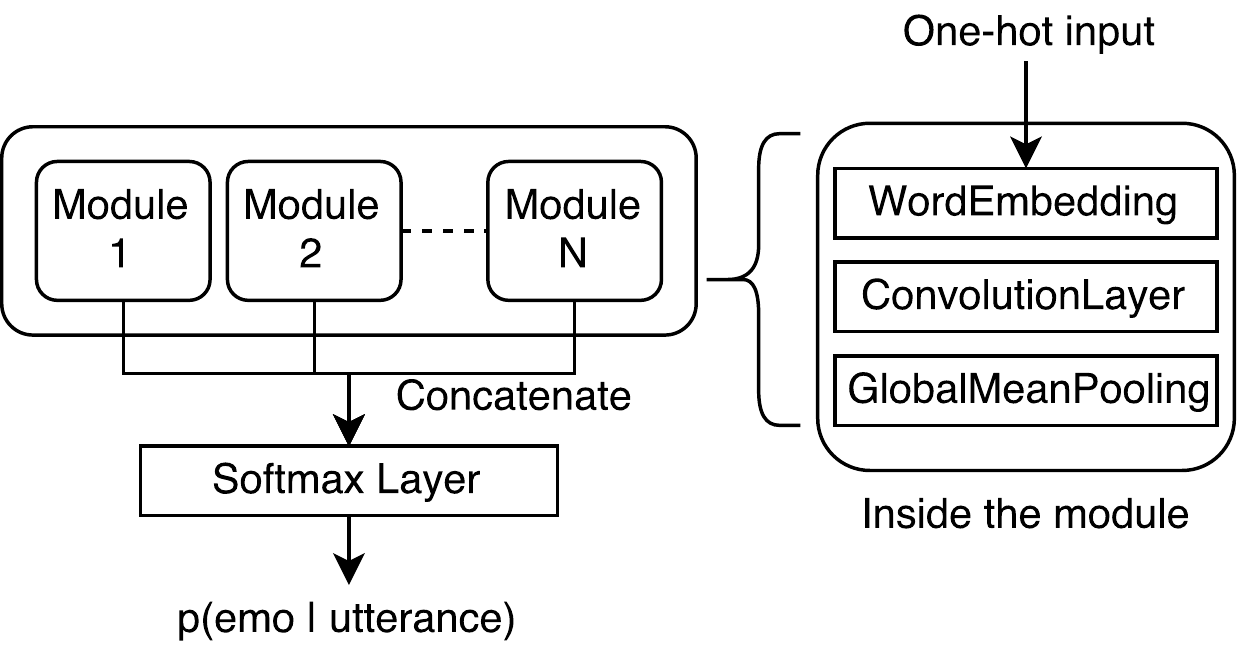}
  \caption{Multi-resolution CNN architecture.}
  \label{fig:CNN}
\end{figure}

This CNN system has multiple parallel modules, each with a convolution layer of 
different kernel size that slides through words in a given utterance to exploit the textual information at different context ranges. 
Each module has a global average pooling layer in the end to aggregate the hidden representations of different words into a unique utterance-level embedding. 
The embeddings over all parallel modules were concatenated to obtain the final utterance embedding. Finally, a fully connected layer with softmax activation
computes the emotion classes' posteriors. 
The detailed components are shown in Figure~\ref{fig:CNN}. 
The number of parallel modules $N$ in the figure and the kernel size for each of them were determined differently for each dataset. 
We explain how they were decided in Section 5.

This network was trained with a combination of classification and verification objectives. On the one hand, cross-entropy objective
on the predicted posteriors was used to improve the classification accuracy. On the other hand, we used a siamese network~\cite{Siamese_YannLecun} in which
the embeddings of two utterances are compared. We use binary-cross entropy loss to make embeddings for same emotion utterances closer and to make 
embeddings from different emotions separate. This is similar to a verification task like speaker verification~\cite{snyder-xvec}. To compare embeddings, we used
cosine scoring followed by a sigmoid function.
\begin{equation} \label{cosine_similarity}
p_{\mathrm{A-B}}=\frac{1}{1+e^{-\cos(d(A),d(B))}} \;.
\end{equation}
where $d(\mathrm{A})$ and $d(\mathrm{B})$ are the embeddings of utterance A and B.

Then, the binary cross entropy for verification loss is
\begin{equation}
	\resizebox{0.91\hsize}{!}{%
$V(A,B) = -t_{\mathrm{A-B}}\log(p_{\mathrm{A-B}}) - (1-t_{\mathrm{A-B}})\log(1-p_{\mathrm{A-B}})$%
	}
\end{equation}
where $t_{\mathrm{A-B}}$ is a target label that becomes 1 if utterance A and B  are from same emotion and 0 otherwise.

In the end, we optimize the objective,
\begin{equation} \label{Total_Obj}
C = \sum_{\mathrm{A}}\sum_{\mathrm{B} \ne \mathrm{A}} H_\mathrm{A} + \lambda V(\mathrm{A},\mathrm{B})
\end{equation}
where $H_\mathrm{A}$ is categorical cross entropy and $\lambda$ is a scale factor to balance the weight of classification and verification objectives; 
the sum is calculated over all possible utterance pairs within mini-batch.

In this paper, emotion vector (e-vector) is compared with this proposed CNN system. The e-vector is a $D$ dimensional feature vector calculated from the equations in~\cite{jin2015speech}.
Here, $D$ corresponds to the number of classes. The value of each element in the vector is the average of all the words' weights in an utterance where the weights indicate its inclination for a specific emotion. The classification based on e-vector showed the highest accuracy among all the single systems proposed in~\cite{jin2015speech}.

\section{Database}

\subsection{IEMOCAP}
USC-IEMOCAP~\cite{busso2008iemocap}
is a database where two actors communicate in each session to elicit specific type of emotions.
It consists of 5 sessions acted by 10 different professional actors.
They either perform selected emotional scripts or improvise hypothetical scenarios.
The recorded videos are annotated by human annotators.
Since labeling emotions can be subjective depending on person,
each utterance was annotated by at least three annotators for categorical labels such as
angry and happy or for dimensional labels such as valence and activation.
For the experiments in this paper,
we used only utterances annotated as one of the following categorical emotion labels:
angry, happy, excited, sad, and neutral.
Besides, we only used recordings where majority of annotators agreed on the emotion labels.
\emph{Happy} and \emph{excited} emotions were combined as \emph{happy}
in order to balance the number of samples in each emotion class.
In the end, we had 1103 utterances for angry, 1636 for happy, 1084 for sad, and 1708 for neutral that sum up to 5531 in total.
\footnote{Refer to the last page}
\st{We ran leave-one-speaker-out 10-fold cross validation where 8 folds were used for training, one for validation, and one for evaluation respectively.}
\subsection{Call center data}
This dataset consists of 1842 telephone calls from call centers.
The calls were segmented into utterances using ASR and they were annotated utterance-wise with three emotion labels:
negative, positive or neutral.
After eliminating cross-talk, silence, and noisy parts, we obtained 5160 utterances for negative, 1735 for positive, and 161898 for neutral.
Due to its nature, most of the utterances were labeled as neutral.
Neutral and negative utterances were randomly sampled to balance the number of data labeled as positive.
We experimented using 5-fold cross-validation where there was no customer overlap between folds.

While for IEMOCAP we used human annotated transcripts, here we generated transcripts using Kaldi ASR system.
The acoustic models for speech recognition were trained using Fisher and Switchboard data-sets. The final speech recognition engine used voice activity detection to remove long silences and segment the audio for processing.
The language models were subsequently replaced with in-domain language models. We used these ASR transcripts in our experiments without any post-processing except for removing special characters.

\begin{figure}
  \centering
  \includegraphics[width=\linewidth]{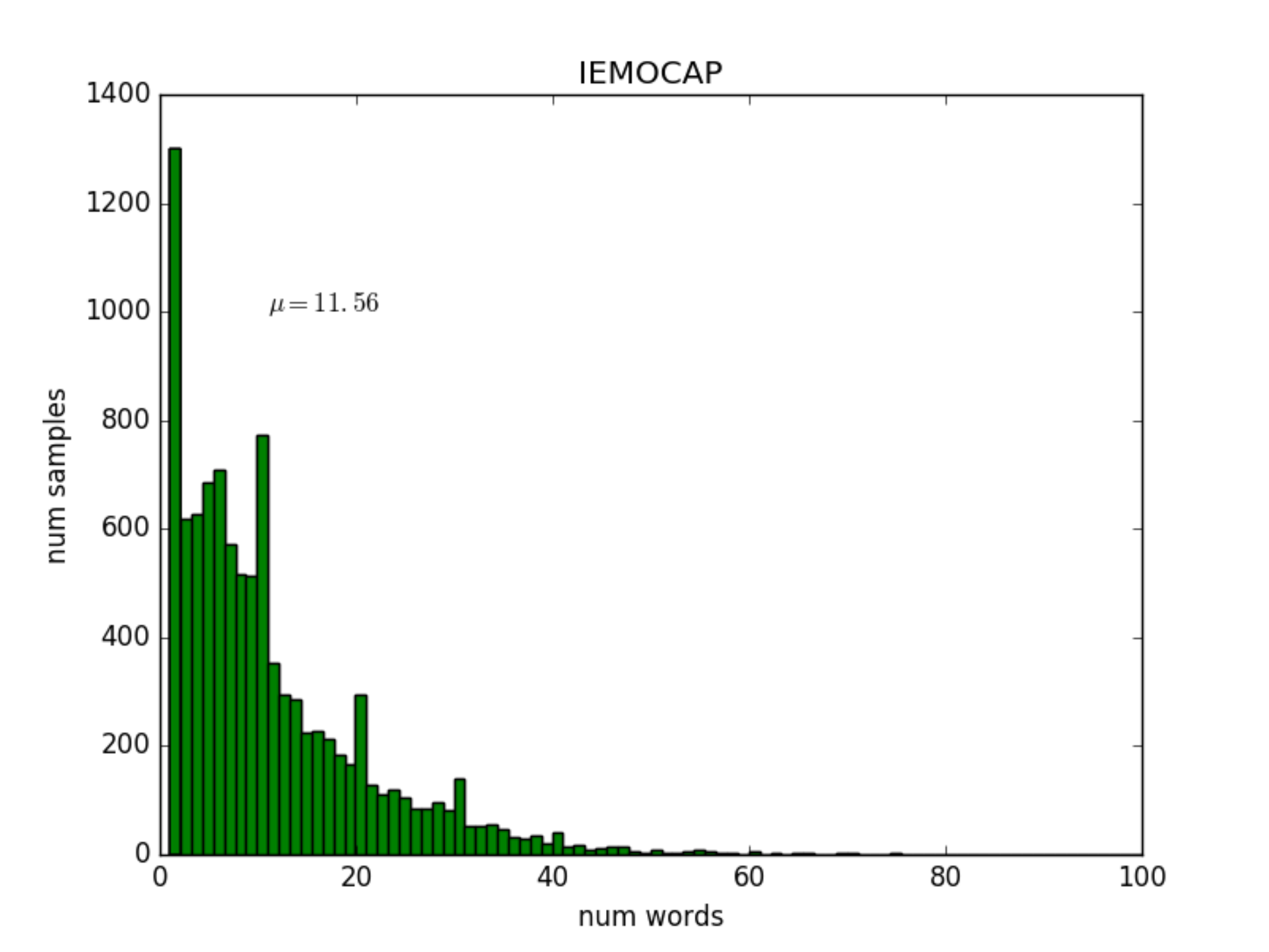}
  \caption{histogram of word length in IEMOCAP }
  \label{fig2}
\end{figure}

\begin{figure}
  \centering
  \includegraphics[width=0.9\linewidth]{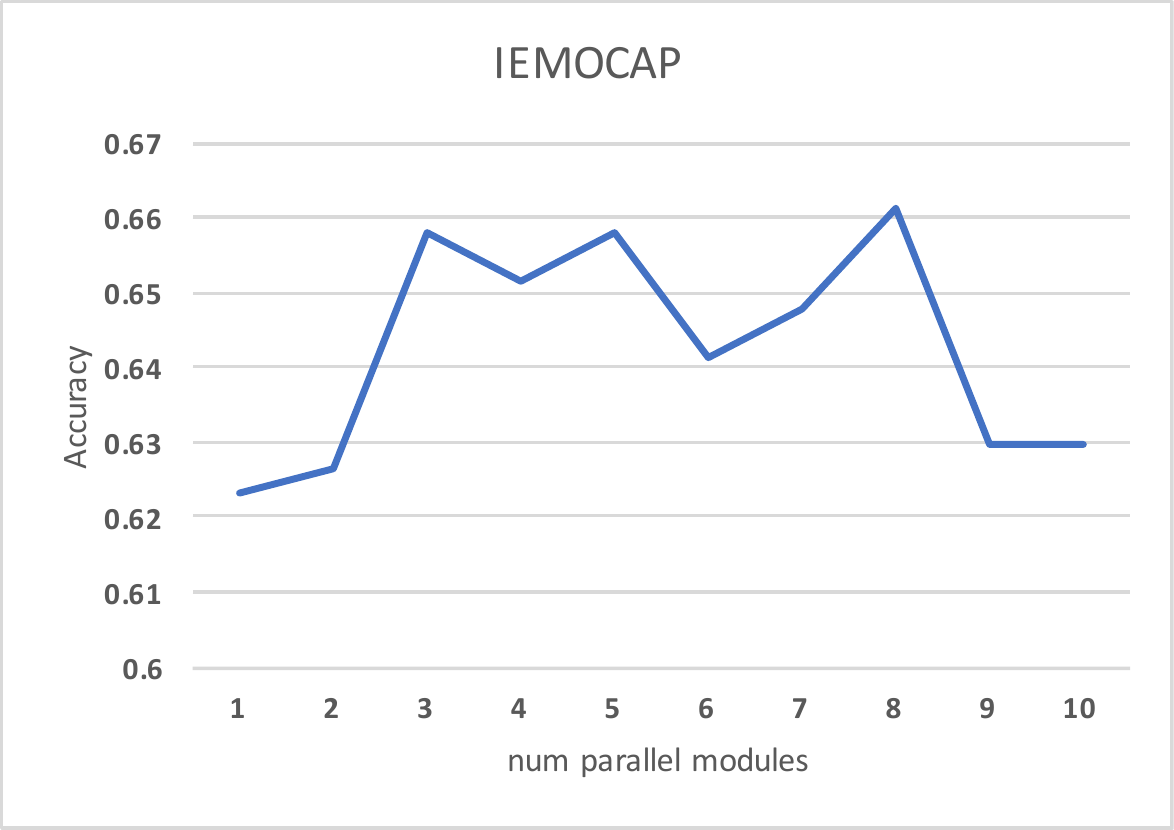}
  \caption{Accuracy vs the number of modules}
  \label{fig3}
\end{figure}

\section{Experiments}

\subsection{Experimental setup}
Through this section, we mainly explain the set up for the proposed LSTM and CNN. After experiments of each system individually, the fusion system composed of their combinations were also explored. When we fuse the systems, we simply concatenate the scores from each system or e-vector (e.g. outputs of softmax layers in case of LSTM and CNN or e-vector itself.) and feed them into a following SVM to predict final emotion labels.

\subsubsection{IEMOCAP}

First, we experiment with the multi-resolution CNN IEMOCAP human-annotated transcripts.
Figure~\ref{fig2} shows statistics of number of words per utterance, which is 11.56 on average.

The multi-resolution CNN used 4 modules with kernel sizes 1, 4, 7, and 11.
The number of modules was selected heuristically.
To select the number, we set the the verification loss weight $\lambda=0$ and experimented
by changing the number of parallel modules from 1 to 10.
The result is shown in Figure~\ref{fig3}.
Accuracy did not improve significantly adding more than three modules.
Since most of the utterances have their length 11 or 12, as seen in Figure~\ref{fig2},
it is a waste for the most of the utterances to set kernel size bigger than 12.
Thus, we finally chose to use four parallel modules having maximum kernel size 11.
The weight for the verification loss $\lambda$ was determined based on the validation set during 10-fold cross validation
ranging from 0.05 to 0.15.

The acoustic feature based LSTM had 2 forward LSTM layers with 256 units per layer.
The hidden representations over time steps were averaged by a
global mean pooling layer, which was followed by 2 dense layers with 256 units and 4 units (the number of classes) respectively.
Dropout with 0.5 drop probability was used for the first dense layer to improve generalization. During the training, adam optimizer with default setting in Keras was used, and batch size was set to 40.

\subsubsection{Call center data}

In experiments using call center data, the average number of words per utterance was 6.73 and the median was 3.
Due to this fact, it was too aggressive to set 3 as kernel size increment between parallel modules.
Thus, we set the increment as 1 for this experiment. Through the grid search, we chose  $\lambda=0.15$ and 3 parallel modules.

For the acoustic LSTM, we used a LSTM layer with only 96 units to avoid over-fitting caused by the limited amount of training samples.
We do not have many emotional utterances per person in this call center data.
Since the data came from 1842 calls meaning roughly 1842 $\times$ 2 (customer + agent) = 3684 people are there,
only one positive utterance is available from more than 2 people on average.

\subsection{Experiment results}
To measure performance of systems, we report overall accuracy on test examples (weighted accuracy, WA) and average recall over different emotion categories (unweighted accuracy, UA) in addition to recall in each class. Notice that we did not report WA for call center data due to its data imbalance over classes.
                                                                                                                                                                                                                                                                                                                                                                                                          
Table~\ref{tab1} presents accuracy results for IEMOCAP dataset. It compares results for several individual systems and the fusion of them.
The results show that the multi-resolution CNN (MCNN) improved by 4\% over hand-crafted e-vector feature, and fusion of both improved by 6\% relative in WA.
Interestingly, the acoustic system is worse than text based systems. However, it contains complementary information and fusion improved the performance significantly.
Fusion of the three systems improved by 21\% relative w.r.t. single acoustic system and by 12\% w.r.t. MCNN in WA.

\begin{table}[h!]
\centering
\caption{Performance (\%) of single systems and their fusion on IEMOCAP dataset}
\label{tab1}
\resizebox{0.47\textwidth}{!}{%
\begin{tabular}{@{}lcccccc@{}}
\hline
System                 & angry & happy & sad   & neutral & WA & UA\\ \hline
LSTM                   & 52.47 & 43.22 & 62.40 & 54.8    & 53.43 & 53.23  \\
E-vector               & 59.57 & \bf 70.6  & 49.98 & 48.84   & 56.22 & 57.25   \\
MCNN                   & 60.72 & 59.16 & 52.94 & \bf 62.39   & 58.47 & 58.80  \\
E-vector + MCNN        & 64.44 & 64.24 & 48.18 & 60.73   & 59.63 & 59.40  \\
MCNN + LSTM            & 70.23 & 64.07 & 65.98 & 57.02   & 63.17 & 64.33  \\
E-vector + MCNN + LSTM & \bf 72.38 & 67.50 & \bf 66.50 & 57.35   & \bf 64.97 & \bf 65.9    \\ \hline                                                                                                                                                                                                                                                                                                                 
\end{tabular}%
}
\end{table}

Table~\ref{tab2} compares
the CNN system and the acoustic and textual fusion system with the systems in~\cite{poria2016MKL}.
The proposed CNN system using textual features is better in predicting angry and neutral while the fusion system of acoustic and textual in this paper outperforms in angry and sad.

\begin{table}[h!]
\centering
\caption{Comparison with previous work}
\label{tab2}
\resizebox{0.45\textwidth}{!}{%
\begin{tabular}{llllll}
\hline
Modality                                                                   &                      & angry & happy & sad   & neutral \\ \hline
\multirow{2}{*}{Text}                                                      & Poria et al.~\cite{poria2016MKL} & 60.01 & 58.71 & 57.15 & 61.25   \\
                                                                           & MCNN                 & 60.72 & 59.16 & 52.94 & 62.39   \\ \hline
\multirow{2}{*}{\begin{tabular}[c]{@{}l@{}}Text +\\ Acoustic\end{tabular}} & Poria et al.~\cite{poria2016MKL} & 62.5  & \bf 65.21 & 63.3  & \bf 69.25   \\                                                                                                                                                                                                                                                            
                                                                           & MCNN + LSTM          & \bf 70.24 & 64.07 & \bf 65.98 & 57.02  \\ \hline
\end{tabular}%
}
\end{table}

Table~\ref{tab3} compares single systems and the fusion systems on call center data. The fusion of MCNN and LSTM compared to a single LSTM showed 3.4\% relative improvement in UA. Considering the fusion of two systems both trained on transcripts (E-vector +MCNN) did not improve at all compared to single E-vector and MCNN systems, the result from the LSTM, MCNN fusion suggests that there is complementary information between acoustic features and transcripts as it was also shown in the previous experiments on IEMOCAP. The fusion of three systems showed best UA although recall per class was not the best. This suggests that more sophisticated fusion method can improve the fusion system better.

\begin{table}[h!]
\centering
\caption{System comparison on call center data}
\label{tab3}
\resizebox{0.45\textwidth}{!}{%
\begin{tabular}{lllll}
\hline
System                 & negative & positive & neutral &  UA \\ \hline
LSTM                   & 42.6     & 46.53    & \bf 39.69   & 42.93    \\
E-vector               & \bf 63.2     & 37.65    & 33.32   & 44.72    \\
MCNN                   & 60.4     & 36.49    & 29.54   & 42.14    \\
E-vector + MCNN & 42.18     & 51.13    & 27.63   & 40.31    \\  
MCNN + LSTM            & 49.38    & \bf 52.90    & 30.92   & 44.40    \\ 
E-vector + MCNN + LSTM & 51.33     & 51.33    & 33.08   & \bf 45.25   \\ \hline
\end{tabular}%
}
\end{table}

\section{Conclusion}

In this paper, we combined information from acoustic features and transcripts to improve emotion recognition from speech where the acoustic system was based on LSTM network.
Meanwhile, a novel multi-resolution CNN (MCNN) was used to predict emotion from ASR transcripts and human annotated transcripts. This MCNN used parallel convolutional layers with
different kernel sizes, which take into account different temporal context ranges. We experimented on the IEMOCAP dataset and call center dataset.
In IEMOCAP, the proposed MCNN improved over state-of-the-art hand-crafted features by 4\% in WA. Fusion of MCNN and acoustic LSTM improved WA by 18\% w.r.t. single acoustic
system. This result proves that there is complementary information between acoustic and transcript features.
Finally, the applicability of the fusion system was confirmed through the experiments with ASR transcripts, which were generated from call center data. Even though the transcripts were generated by ASR, the fusion system using both acoustic features and transcripts still outperformed a single system trained on acoustic features by 3.4\% relatively in UA.

In the future, we plan to explore more effective ways of combining information from different modalities such as effective ensemble of classifiers in~\cite{rozgic2012svmensemble} or effective way of combining features from different modalities in~\cite{poria2016MKL}. Also, we will apply the verification loss similarly to the acoustic system in order to improve the system to the degree a system trained on transcripts performs.

\bibliographystyle{IEEEtran}

\bibliography{template}

\begin{thebibliography}{10}
\providecommand{\url}[1]{#1}
\csname url@samestyle\endcsname
\providecommand{\newblock}{\relax}
\providecommand{\bibinfo}[2]{#2}
\providecommand{\BIBentrySTDinterwordspacing}{\spaceskip=0pt\relax}
\providecommand{\BIBentryALTinterwordstretchfactor}{4}
\providecommand{\BIBentryALTinterwordspacing}{\spaceskip=\fontdimen2\font plus
\BIBentryALTinterwordstretchfactor\fontdimen3\font minus
  \fontdimen4\font\relax}
\providecommand{\BIBforeignlanguage}[2]{{%
\expandafter\ifx\csname l@#1\endcsname\relax
\typeout{** WARNING: IEEEtran.bst: No hyphenation pattern has been}%
\typeout{** loaded for the language `#1'. Using the pattern for}%
\typeout{** the default language instead.}%
\else
\language=\csname l@#1\endcsname
\fi
#2}}
\providecommand{\BIBdecl}{\relax}
\BIBdecl

\bibitem{picard2000affective}
R.~W. Picard, \emph{Affective computing}.\hskip 1em plus 0.5em minus
  0.4em\relax MIT press, 2000.

\bibitem{poria2017review}
S.~Poria, E.~Cambria, R.~Bajpai, and A.~Hussain, ``A review of affective
  computing: From unimodal analysis to multimodal fusion,'' \emph{Information
  Fusion}, vol.~37, pp. 98--125, 2017.

\bibitem{kim2013deep}
Y.~Kim, H.~Lee, and E.~M. Provost, ``Deep learning for robust feature
  generation in audiovisual emotion recognition,'' in \emph{Acoustics, Speech
  and Signal Processing (ICASSP), 2013 IEEE International Conference on}.\hskip
  1em plus 0.5em minus 0.4em\relax IEEE, 2013, pp. 3687--3691.

\bibitem{rozgic2012svmensemble}
V.~Rozgic, S.~Ananthakrishnan, S.~Saleem, R.~Kumar, and R.~Prasad, ``Ensemble
  of svm trees for multimodal emotion recognition,'' in \emph{Signal \&
  Information Processing Association Annual Summit and Conference (APSIPA ASC),
  2012 Asia-Pacific}.\hskip 1em plus 0.5em minus 0.4em\relax IEEE, 2012, pp.
  1--4.

\bibitem{mirsamadi2017rnnattention}
S.~Mirsamadi, E.~Barsoum, and C.~Zhang, ``Automatic speech emotion recognition
  using recurrent neural networks with local attention,'' in \emph{Acoustics,
  Speech and Signal Processing (ICASSP), 2017 IEEE International Conference
  on}.\hskip 1em plus 0.5em minus 0.4em\relax IEEE, 2017, pp. 2227--2231.

\bibitem{poria2017context}
S.~Poria, E.~Cambria, D.~Hazarika, N.~Majumder, A.~Zadeh, and L.-P. Morency,
  ``Context-dependent sentiment analysis in user-generated videos,'' in
  \emph{Proceedings of the 55th Annual Meeting of the Association for
  Computational Linguistics (Volume 1: Long Papers)}, vol.~1, 2017, pp.
  873--883.

\bibitem{jin2015speech}
Q.~Jin, C.~Li, S.~Chen, and H.~Wu, ``Speech emotion recognition with acoustic
  and lexical features,'' in \emph{Acoustics, Speech and Signal Processing
  (ICASSP), 2015 IEEE International Conference on}.\hskip 1em plus 0.5em minus
  0.4em\relax IEEE, 2015, pp. 4749--4753.

\bibitem{raghu2018mcnn}
R.~Pappagari, J.~Villalba, and N.~Dehak, ``Joint verification-identification in
  end-to-end multi-scale cnn framework for topic identification,''
  \emph{proceedings of ICASSP}, 2018.

\bibitem{hochreiter1997lstm}
S.~Hochreiter and J.~Schmidhuber, ``Long short-term memory,'' \emph{Neural
  computation}, vol.~9, no.~8, pp. 1735--1780, 1997.

\bibitem{chung2014gru}
J.~Chung, C.~Gulcehre, K.~Cho, and Y.~Bengio, ``Empirical evaluation of gated
  recurrent neural networks on sequence modeling,'' \emph{arXiv preprint
  arXiv:1412.3555}, 2014.

\bibitem{chollet2015keras}
F.~Chollet \emph{et~al.}, ``Keras,'' \url{https://keras.io}, 2015.

\bibitem{eyben2016gemaps}
F.~Eyben, K.~R. Scherer, B.~W. Schuller, J.~Sundberg, E.~Andr{\'e}, C.~Busso,
  L.~Y. Devillers, J.~Epps, P.~Laukka, S.~S. Narayanan \emph{et~al.}, ``The
  {Geneva} minimalistic acoustic parameter set (gemaps) for voice research and
  affective computing,'' \emph{IEEE Transactions on Affective Computing},
  vol.~7, no.~2, pp. 190--202, 2016.

\bibitem{eyben2010opensmile}
F.~Eyben, M.~W{\"o}llmer, and B.~Schuller, ``Opensmile: the munich versatile
  and fast open-source audio feature extractor,'' in \emph{Proceedings of the
  18th ACM international conference on Multimedia}.\hskip 1em plus 0.5em minus
  0.4em\relax ACM, 2010, pp. 1459--1462.

\bibitem{Siamese_YannLecun}
S.~Chopra, R.~Hadsell, and Y.~LeCun, ``Learning a similarity metric
  discriminatively, with application to face verification,'' in \emph{Computer
  Vision and Pattern Recognition, 2005. CVPR 2005. IEEE Computer Society
  Conference on}, vol.~1.\hskip 1em plus 0.5em minus 0.4em\relax IEEE, 2005,
  pp. 539--546.

\bibitem{snyder-xvec}
D.~Snyder, P.~Ghahremani, D.~Povey, D.~Garcia-Romero, Y.~Carmiel, and
  S.~Khudanpur, ``{Deep neural network-based speaker embeddings for end-to-end
  speaker verification},'' in \emph{Proceedings of the 2016 IEEE Spoken
  Language Technology Workshop (SLT)}.\hskip 1em plus 0.5em minus 0.4em\relax
  San Diego, CA, USA: IEEE, dec 2016, pp. 165--170.

\bibitem{busso2008iemocap}
C.~Busso, M.~Bulut, C.-C. Lee, A.~Kazemzadeh, E.~Mower, S.~Kim, J.~N. Chang,
  S.~Lee, and S.~S. Narayanan, ``Iemocap: Interactive emotional dyadic motion
  capture database,'' \emph{Language resources and evaluation}, vol.~42, no.~4,
  p. 335, 2008.

\bibitem{poria2016MKL}
S.~Poria, I.~Chaturvedi, E.~Cambria, and A.~Hussain, ``Convolutional {MKL}
  based multimodal emotion recognition and sentiment analysis,'' in \emph{Data
  Mining (ICDM), 2016 IEEE 16th International Conference on}.\hskip 1em plus
  0.5em minus 0.4em\relax IEEE, 2016, pp. 439--448.

\end{thebibliography}

\newpage
\textbf{Note}

\textsuperscript{1}It was not caught when published in Interspeech 2018 but found later that each fold was composed of a session where an actor wears markers in the session. For example, \textit{Ses01F*} utterances and \textit{Ses01M*} ones separate into two different folds where \textit{Ses01} and \textit{F/M} mean session index and who wears markers respectively. Regardless of who wears markers in a fold, there are still two actors in the fold. Thus, there could be speaker overlap between some folds. Sincere apology I could not find this before published and cited by some people. Other than that, however, all others including gain/loss from implemented systems reported in this paper are correct and remain the same as published in Interspeech 2018.

\end{document}